\documentclass{article}

\usepackage{arxiv}
\usepackage[utf8]{inputenc} 
\usepackage[T1]{fontenc}    
\usepackage{url}            
\usepackage{booktabs}       
\usepackage{amsfonts}       
\usepackage{nicefrac}       
\usepackage{microtype}      
\usepackage{graphicx}
\usepackage{doi}

\hypersetup{
pdftitle={Human-agent coordination in a group formation game},
pdfsubject={physics.soc-ph, cs.GT},
pdfauthor={Tuomas Takko, Kunal Bhattacharya, Daniel Monsivais, Kimmo Kaski},
pdfkeywords={collective intelligence, coordination game, human-agent hybrids},
}

\title{Human-agent coordination in a group formation game}

\author{Tuomas Takko \\
Department of Computer Science\\
Aalto University School of Science\\
00076, Finland \\
\texttt{tuomas.takko@aalto.fi}
\And
Kunal Bhattacharya \\
Department of Industrial Engineering and Management\\
Department of Computer Science\\
Aalto University School of Science\\
00076, Finland \\
\And
Daniel Monsivais \\
Department of Computer Science\\
Aalto University School of Science\\
00076, Finland \\
\And
Kimmo Kaski \\
Department of Computer Science\\
Aalto University School of Science\\
00076, Finland \\
The Alan Turing Institute\\
96 Euston Rd, Kings Cross, London NW1 2DB, UK
}

\begin{document}
\maketitle

\begin{abstract}
Coordination and cooperation between humans and autonomous agents in cooperative games raises interesting questions of human decision making and behaviour changes. 
Here we report our findings from a group formation game in a small-world network of different mixes of human and agent players, aiming to achieve connected clusters of the same colour by swapping places with neighbouring players using non-overlapping information. 
In the experiments the human players are incentivized by rewarding to prioritize their own cluster while the model of agents' decision making is derived from our previous experiment of purely cooperative game between human players. 
The experiments were performed by grouping the players in three different setups to investigate the overall effect of having cooperative autonomous agents within teams.         
We observe that the change in the behavior of human subjects adjusts to playing with autonomous agents by being less risk averse, while keeping the overall performance efficient by splitting the behaviour into selfish and cooperative in the two actions performed during the rounds of the game. 
Moreover, results from two hybrid human-agent setups suggest that the group composition affects the evolution of clusters. 
Our findings indicate that in purely or lesser cooperative settings, providing more control to humans could help in maximizing the overall performance of hybrid systems.

\end{abstract}

\thispagestyle{empty}

\noindent

\section*{Introduction}
In digitized human societies interactions between humans and artificial autonomous agents is becoming more and more commonplace.
This calls for deeper understanding and study of the possible outcomes of these interactions in different social situations or setups, e.g. in games, healthcare, retail stores and transportation \cite{kearns2006experimental,jennings2014human,isern2010agents, corchado2008intelligent, van2017domo,robu2013online}.   
Given the fact that human behaviour is far from homogeneous, it may be difficult to predict the emergent behaviour in these systems. 
If one wants to realize a system, where the desired outcomes include cooperation or coordination between the humans and agents, understanding of the macro-level dynamics in terms of the different types of micro-level human-human, human-agent and agent-agent interactions, is needed. 
In addition, there is need for benchmarking models to take into account the variability in human psychological preferences that influence their decision making~ \cite{bonabeau2002agent,kahneman2003maps,groom2007can, bhattacharya2019group}.

Games and online games in particular provide valuable frameworks for studying dynamics of human-agent collectives as well as purely human groups. 
As an example, human-agent games that require cooperation from the human subjects, such as in the case of iterated Prisoner's Dilemma, have been studied with regards to the human volunteer's perception on the opposite player being an agent or a human~\cite{ishowo2019behavioural}.
In this study, an interesting finding is that if the volunteers were told that they were playing with an agent, the level of cooperation was found to decrease. 
Similarly, introducing social networks into the design of studies have shed light on the non-local influence of the interactions taking place in games. 
For instance, Shirado and Christakis \cite{shirado2017locally} studied how the collective performance of humans trying to solve a coordination game on a network changes in the presence of agents (or bots), and showed the impact of the degree of randomness in agents' behaviour on the outcome of the game.
In other words, the inclusion of randomly acting autonomous agents was found to increase the overall performance of group consisting of humans and agents. 
Also, games with network-based systems of agents with heterogeneous behaviour have been studied in the context of cooperation and evolutionary behaviour. \cite{huang2018heterogeneous, huang2018incorporating, perc2007cyclical}

In this paper, we investigate a hybrid system of humans and autonomous agents in a cooperative game setup played on a virtual network. 
In order to observe how the inclusion of cooperative autonomous agents affects the outcomes, we use a model fitted on observed human behaviour from our prior experiments~\cite{bhattacharya2019group}. 
In two cases we set up the experiments such that the agents are distributed into groups of humans in varying proportions and the resulting dynamics are compared to two control experiments, one with only humans and another with only agents. 
The framework used in the present study and in our previous research \cite{bhattacharya2019group} is in the spirit of the earlier works by Kearns and colleagues (see \cite{kearns2012behavioral} and the references therein). 
In their work they focused on studying the effect of network structure on the efficiency of solving problems like the graph coloring and consensus by human subjects \cite{kearns2012behavioral,kearns2006experimental,judd2010behavioral}. 
The game of group formation also shares similarities to the matching problem that considers members of two distinct sets forming pairs for their own mutual benefit \cite{gale1962college,laureti2003matching}. 

In the context of experiments involving problem-solving tasks the complex relationship between the individual-level human behaviour, collective performance, and network properties, has been explored in several studies \cite{baronchelli2010modeling,guazzini2015modeling, centola2015spontaneous}. 
It has been shown that coordination, cooperation, and other social actions within human groups can be described and analyzed through carefully designed online and incentivized experiments.
In our experimental setup the players are incentivized to arrange themselves in groups. 
This game can be placed in the context of game-theoretic studies of social group formation, for example, such as the games in \cite{chauhan2018Schelling,elkind2019schelling} that are based on Schelling's segregation model \cite{schelling1971dynamic}, and more generally, the hedonic coalition formation games \cite{dreze1980hedonic,aziz2016hedonic}.

Our overall approach, includes two sets of experiments, a previous work~\cite{bhattacharya2019group} and the current study. 
In the former we conducted an online computer lab-based experiments with human subjects who were incentivized to form connected clusters or groups on a small-world network by coordinating and cooperating with members of other teams with non-overlapping information. 
From the results of the previous study we constructed a data-driven model of an autonomous agent to replicate the decision making of the `cooperating' subjects. 
In the current setup we combine the humans and autonomous agents in games (detailed below) with a different incentive scheme for the humans, while for the agents we use the model of humans from the previous experiment. 
Our broad motivation is to gain insight into the effects of including cooperative autonomous agents in a game, where a certain degree of competition is allowed between teams that includes human subjects. 
We note that, like in an earlier study of a hybrid human-agent game \cite{shirado2017locally}, in our hybrid games the human players were not informed to be playing with human-like agents. 
While the teams are allowed to work towards the goal to maximize their own benefit we examine whether the overall benefit can be maximized along the lines of public goods~\cite{barrett2007cooperate}.

\section*{Materials and Methods}
\begin{figure}[htb]
\centering
\includegraphics[width=1.0\linewidth]{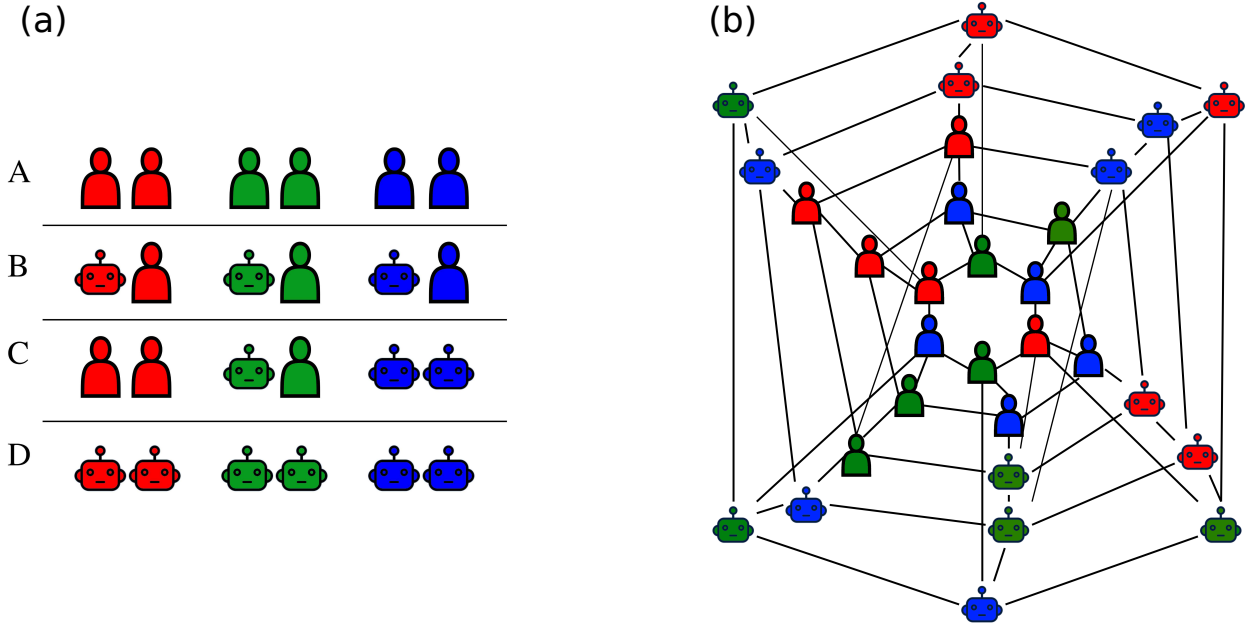}
\caption{\textbf{(a) Composition of the groups in experimental setups.}
Each icon represents 5 players or autonomous agents or bots. 
Setup A consisted of only human players, the hybrid setups B and C had 15 human and 15 agent players with varying group consistencies, and setup D consisted of only agent players. 
The difference in group consistencies between setups B and C provided different numbers of human-human, human-bot and bot-bot interactions.
\textbf{(b) An example snapshot from a game with setup B.}
The initial configuration of a game with setup B. The largest clusters are red 7, green 4 and blue 2.
Each network configuration was randomized with the restriction that none of the groups could have more than 6 points at the start of each game and that none of the nodes would have degree above 5.
}
\label{treatmentStructure}
\end{figure}

The game of group formation is played on a fixed and regular network of $n$ nodes ~\cite{bhattacharya2019group}. 
The players are divided into $m$ equal sized groups, where each group is identified by a colour, and the number of players in each group is $l$, such that, $n={m}\times{l}$. 
The general objective for the players of the game is to maximize their group's cluster size by exchanging positions in the network, such that all the players in a group eventually form a connected cluster.
An example of the network configuration and the clusters in it is illustrated in Fig.~\ref{treatmentStructure} (right). 
The game is played in rounds where on a given round the players of one colour send requests to the players of a different colour located on the neighbouring nodes in order to swap places.
These requests are then either accepted or rejected by the players on the receiving nodes.
The colour for which it is turn to send a request is changed cyclically such that each colour has the same amount of opportunities to make requests.
The maximum number of rounds that can be played in each game is $r$. Thus, each player would have at most $r/m$ opportunities to send a request during a game if the latter is not completed before the $r$-th round. Similarly, each player would have  maximally $2r/m$ opportunities to receive requests. In a given round, nodes having a colour different from the requesting nodes can simultaneously receive multiple requests. The actual number of opportunities to interact during a game depends also on the position of the player in the network. For example, players with all the neighbours belonging to its own group can not send or receive requests. 
To request neighbours with the same color is forbidden, because such an exchange of places would not benefit the the overall objective of the game or change the state of the game.
During the course of the game the amount of information provided to the players (both humans and  autonomous agents) is limited to the local neighbourhood in the network, the current number of points the player's group has, and the global information about the largest clusters of each colour. 
The local information provided about the neighbourhood consists of the colour and the cluster size of the players in the nodes that are directly linked to the player. 
Nodes and edges that are not in the immediate neighbourhood of the player are displayed in grey colour (masked). 
The game is terminated once the designated objective is achieved or when a certain number of rounds is reached.

In our previous work \cite{bhattacharya2019group}, the recruited human subjects played the game of group formation in order to achieve a cooperative payoff.
The players were expected to collectively achieve an overall configuration on the network such that all the $m$ groups (colours) would finally attain the maximum cluster size ($l$).
The experiment consisted of $m=3$ coloured groups each with $l=10$ human players, placed on a regular network with periodic boundary and randomized small-world links similar to Fig.~\ref{treatmentStructure} (right). 
In the game the payoff function was based on collective progress ($ACP$), measured by calculating the average of the normalized size of the three largest clusters for each colour. 
The game was concluded once the players reached the maximized clusters ($ACP = 1.0$) or the number of rounds played reached $21$. 
To facilitate a larger number of exchanges in the game, the initial setup of the network was in a graph coloured formation, where the average collective progress would be $0.1$, after which we added random small-world links between nodes with the restriction that the degree would not exceed 5.

In order to understand and quantify human behavior during the games, we implemented and trained a probability matching based model by utilizing the local information of the player's neighbourhood including the cluster sizes of the neighbours and their respective colours.
Furthermore, we evaluated the level of rationality and perception of risk, the human subjects were showing in the experimental sessions using this model and the obtained parameters.
It was found that the the human players were successful in forming the maximum clusters in most of the games of the experimental session.  
The results obtained by varying the model's parameter weights during simulations suggest that the decision making and utilization of the provided limited information was effective and the perception of risk was close to optimal when the objective was purely cooperative.

\begin{table}[htb]
\caption{The rewarding scheme applied in the experiment. 
The players obtained points according to the requirements in the table and lost the respective amount of points if the requirement was no longer satisfied. With this scheme the maximum amount of points in a single game was 20.\cite{takko2019study}\label{pointTable1}}
\begin{center}
 \resizebox{0.55\textwidth}{!}{
\renewcommand{\arraystretch}{1.5}
\begin{tabular}{|l|c|}
\hline
\textbf{Requirement}                                           & \textbf{Points} \\ \hline
\hline
The player's respective group has cluster size of 6 or greater & 5                                   \\
The  player's respective group has cluster size 9 or 10        & 7                                   \\
One of the other groups reaches cluster size of 9 or 10        & 4    \\           
\hline
\end{tabular}
}
\end{center}
\end{table}

The experimental sessions of the present study were held in a computer lab at Aalto University's campus with 30 volunteers, recruited via advertisements on social media.
Informed consent was obtained from every volunteer before the experiment with signed consent forms. 
All the volunteers were non-minors. 
The experiment was conducted according to the relevant guidelines and the procedure was approved in advance by the research ethics committee of Aalto University (2017\_02\_BSEN Experiment).
The practical setup was similar to the previous experiment \cite{bhattacharya2019group} in terms of using oTree \cite{otree} as the framework for running the game and restricting the players from mutual communication and the view of others' workstations. 
A view of the graphical user interface can be found in our previous study \cite{bhattacharya2019group}.
The players were introduced to the game by providing them with a presentation and a short tutorial of the game before the experiment started.
The session lasted for four hours, during which a total of 13 games were played. 
A single game consisted of a maximum of 15 rounds and lasted approximately 20 minutes. 
Structurally the experiments were split into sessions of 3 games, which were incentivized with a participation bonus and a reward according to the player's performance. 
Similarly to our previous experiment, the reward was given in the form of movie tickets. 
The show-up bonus for a 3-game session was one movie ticket and the reward was a second movie ticket for gaining a sum of 27 or more points. 
The payoff function for receiving points differs from the one used previously \cite{bhattacharya2019group}. 
Instead of the collective progress to form the maximum clusters, the players were tasked to obtain points by forming larger clusters with the formula illustrated in Table \ref{pointTable1}.
This formula was aimed to direct the players towards a more selfish strategy in comparison to the purely collective objective used in the previous experiment. \cite{takko2019study}

The game was terminated once the maximum payoff for a single game was reached, i.e. the players had formed clusters of size $l$ (points $=20$), or when the initially set number of rounds ($r=15$) was reached. 
The initial network configurations, including the player colours and the added small-world links, were chosen before the experimental session in such a way that the starting configuration would have some degree of clustering whilst being far from completion. 
If the players managed to achieve the required 27 points for the maximum payoff in the first two games of the 3-game session, the third game would not be played as the full incentive for that particular session would already have been achieved. 
These parameters were motivated by the use of simulations with a model (provided below) and by our previous experimental study \cite{bhattacharya2019group}.
The simulations suggested that completing the objective was achievable within the stipulated number of rounds.
Also we note that a less number of rounds ensured the running of more games during the experimental session, thus yielding more data from the different stages of the game such as the initial formation of small clusters.

To investigate human behaviour and decision making in different combinations of human and autonomous agent players we use three experimental setups, A, B, and C. 
The setup A consisted of purely human players ($n=30$) and it served as a baseline for the current rewarding scheme and for the population of players present in the experimental sessions. 
The setups B and C consisted of human-agent collectives with 15 human subjects and 15 autonomous agents, but with different concentrations of humans and agents in each colour group (See Fig. \ref{treatmentStructure}). 
These different group structures were chosen for discovering the possible differences between the number of human-agent interactions and changes in behavior and performance during such hybrid setups. 
The setup B consisted of 3 mixed groups of human subjects and agents in an even ratio (5 human subjects and 5 agents per group). 
The setup C consisted of a purely human group (10 human subjects), a mixed group of humans and agents (5 humans and 5 agents) and a group consisting of  only agents (10 agents). 
In addition to the experimental setups with human subjects we simulated 100 realizations of the game with 30 autonomous agents, which is named setup D. 

In the setups B, C, and D the design of autonomous agents, was obtained from fitting the data from our earlier experimental session to the following model which we describe in brief~\cite{bhattacharya2019group}. 
We assumed that while  requesting or accepting requests, players associate a utility, $P_\omega$ with the choice of a neighbour $\omega$ (the goal being collective). Using probability matching, the probability of choosing the option is
\begin{equation}
p_\omega=\frac{P_\omega/P_0}{1+\sum_{\omega'}P_{\omega'} /P_0}, 
\label{eqn:model}
\end{equation}
where $P_0$ is the utility for not sending a request (that is, not moving). 
A further simplification was done by choosing $P_\omega / P_0=\exp\{\lambda+\alpha s_i+\beta s_j+\delta\langle s(c_j) \rangle\}$ which allowed fitting our experimental data to a logistic function. 
In the last expression, $s_i$ is the cluster size of the focal player, $s_j$ is the cluster size of chosen neighbour and $\langle s(c_j)\rangle$ the average cluster size of rest of the neighbouring players sharing the colour of the chosen neighbour (also see Eq.~\ref{eq:eqn-uj}).
The parameters $\alpha$, $\beta$, $\delta$, and $\lambda$ were obtained from the fit. 
In the current context using this model allows us to mix agents with purely cooperative strategies with humans players.

\begin{figure}[htb]
\centering
\includegraphics[width=0.6\linewidth]{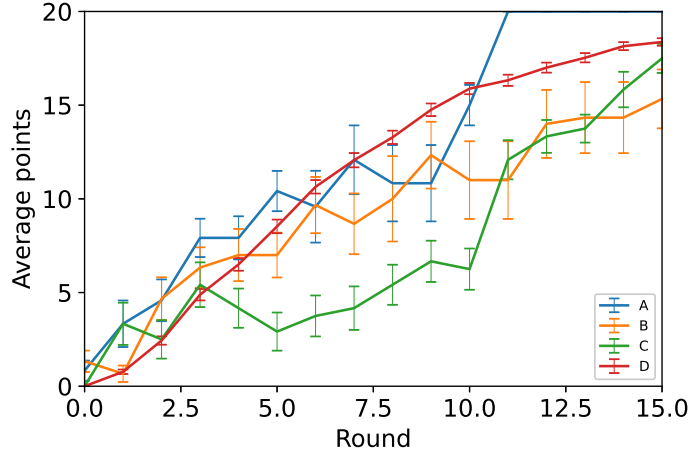}
\caption{\textbf{Average number of points in each round for the different setups.}  
Games with setup A reached the maximum number of points in all of the 4 games while the hybrid setups B and C as well as the setup D did not reach that within the maximum of 15 rounds
The scheme of receiving the maximum reward for each 3-game series resulted in the players playing four games of setup A, five games of setup B and four games of setup C. 
The setup D consists of 100 realizations of the game simulated with cooperative autonomous agents.
Out of the four setups, the hybrid setup B resulted in the worst performance.
However, the difference between the hybrid setups is minor as individual players in one of the colour groups in setup B achieved the maximum reward before the fifth game (i.e. achieved a total sum of 27 points).
The error bars represent the standard errors.
}
\label{pointsRound}
\end{figure}

\begin{figure}[!htb]
\centering
\includegraphics[width=0.7\linewidth]{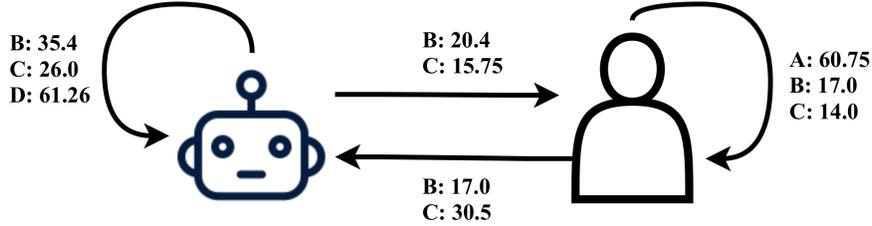}
\caption{\textbf{Average number of interactions between agent and human players of different colour for a game in the experimental setups.} 
The varying agent concentrations in the colour groups resulted in different numbers of interactions between humans and agents. The numbers represent all the initiated interactions (i.e. requests sent). 
The number of initiated interactions in the purely human setup A and the purely agent based setup D are of the same range.
However, the different group compositions in setups B and C have an effect on the number of initiated interactions in terms of human to agent and agent to human. 
It should be noted that the performance of individual groups in the game can lead to situations where a single group has not been able to form a large cluster and the other groups initiate more interactions towards them in order to maximize their own payoff, thus adjusting their behaviour towards more cooperative interactions.
This type of helping behaviour towards the autonomous agents occurred because of the smaller cluster size resulting from the cooperative interactions' tendency to break clusters in order to facilitate movement of other players.
In total, the hybrid setups had more interactions on average (B with 90 and C with 86) than the purely agent or human setups (both A and D with 61). 
}
\label{interactions}
\end{figure}

\begin{figure}[!htb]
\includegraphics[width=1.0\linewidth]{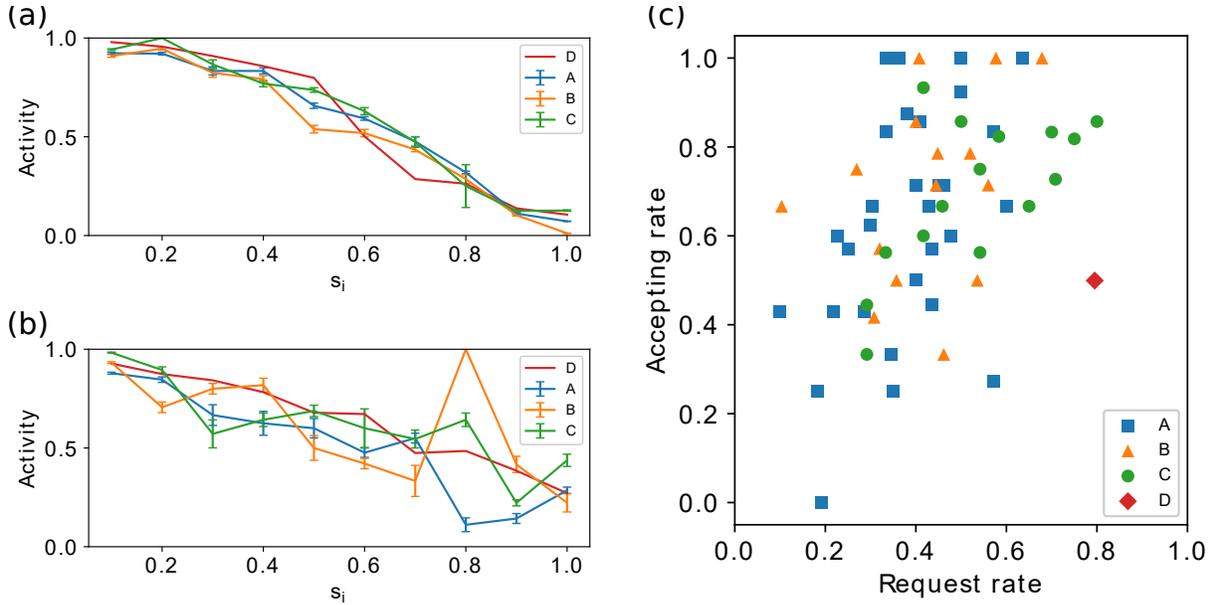}
\caption{\textbf{The requesting (a) and accepting (b) activity of the human players by cluster size in different setups and the autonomous agents' activity in setup D.} 
All of the setups show a similar decay in activity, but it is notable that the players showed a more relaxed accepting activity in the setups with autonomous agents than in the setup with only human players.
This behaviour can be result of the point based rewarding scheme as well as the performance of autonomous agents and their fully cooperative behaviour not resulting in sufficiently large clusters in the hybrid setups.
As the autonomous agents were based on the same model, their activity is also represented in the setup D.
The points are connected for visualizing the trends and the error bars represent standard error of the mean.
\textbf{(c) Overall activity per player in the experimental setups.} 
Each marker represents the fraction of times when a player performed an action having the opportunity to request or accept in a particular setup. 
The distribution of the markers shows that none of the human players requested a swap in every opportunity they had. 
It is notable that in a few instances in setups A and B, the player accepted every incoming request. 
\label{fig_activity}} 
\end{figure}

\begin{figure}[!htb]
\includegraphics[width=1.0\linewidth]{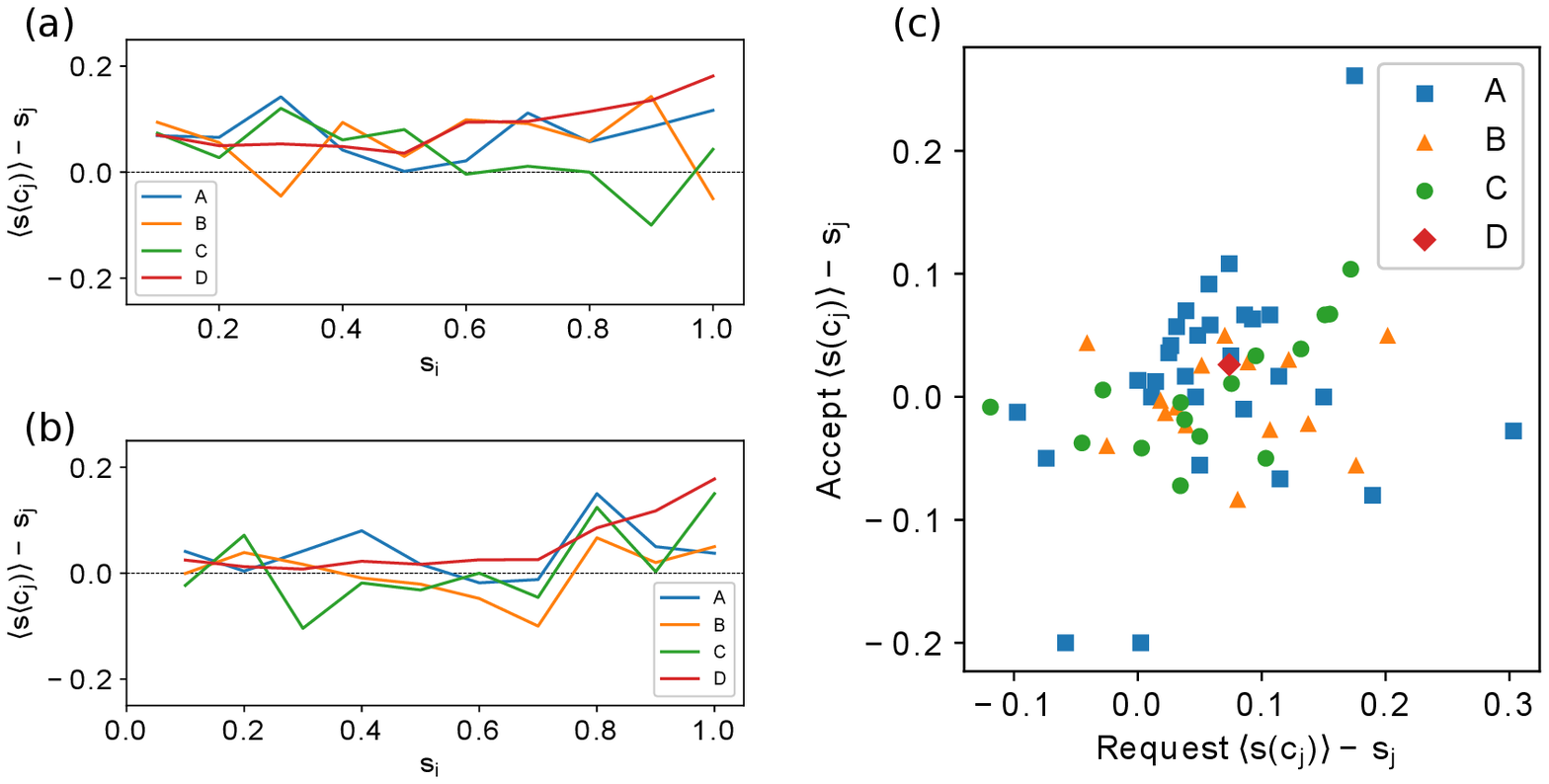}
\caption{\textbf{Requesting (a) and accepting (b) strategies in the experimental setups by cluster size.}
The values are averages of the human interactions in terms of sent requests and accepted requests, where each value is the difference between the normalized average cluster size of the neighbours sharing the chosen neighbour's colour ($\langle s(c_j) \rangle$) and the normalized cluster size of the chosen neighbour ($s_j$). 
Positive values indicate more cooperative decision making in terms of choosing the neighbour with the lowest cluster size of that particular colour and negative values indicate the opposite. 
If the decision making was completely random, the value would be 0. 
\textbf{(c) Individual average strategies in the games.} Each marker represents a player in a particular setup.
The autonomous agents are represented by the average value for all uniformly constructed agents in setup D.
Negative values indicate that the player's strategy on average was not to request the most beneficial neighbour, or accept requests from the neighbour with the smallest cluster size.
\label{strategies}}
\end{figure}

\section*{Results}

During the experiment of the present study a total of two sessions with four games of setup A, two sessions with five games of setup B and two sessions with four games of setup C were played, making a total of 13 games. 
These 13 games resulted in 591 requests sent by the human subjects and 549 requests received by the human subjects.
The experiment started with the group of 30 human subjects playing four games of setup A, after which the pool of human subjects was split into two sets of 15 players each. These two sets of players then separately participated in the hybrid setups B and C, respectively, where the bots were included into the teams. This design enabled us to maintain the network size of 30 in each of the setups (15 humans $+$ 15 bots; see Fig. \ref{treatmentStructure}) and allowed for a comparison between the three setups.
The split into human-agent groups was not disclosed to the players as they proceeded to the next games after the first four games of setup A. 
The agents appeared in the game indistinguishable from the point of view of human subjects, and the fact that the players were now playing with agents was not announced until the end of the experimental session, was intended to reduce a possible bias in the decision making of the human subjects and in order to evaluate ``organic'' adaptation rather than to act as an intervention~\cite{crandall2018cooperating, shirado2017locally}.
The human subjects were effective in reaching the desired final configuration and the maximum obtainable points in each game of setup A, before the end of the given number of rounds (see Fig. \ref{pointsRound}).

The performance of the teams in obtaining points was found to be weaker in the two hybrid setups, especially in the setup B, which had a 3-game session going to the last game before the maximum payoff was obtained for all of the players. 
This difference between the purely human setup A and hybrid setups B and C hints at the human subjects' adaptation to the autonomous agents' decision making was not optimal or alternatively the strategy of the autonomous agents was not optimal for reaching the objective in the given payoff function when playing with human subjects. 
This incompatibility can be the result of the objective for the autonomous agents differing from the current objective of the game due to the model being fit to the data from purely collective games. 
The games in setup B showed a significantly lower performance due to the larger number of agent-agent interactions emerging from the difference in the compositions of the agents and humans in the colour groups (see Fig. \ref{interactions}). 
It should be noted that our autonomous agent model from our earlier study \cite{bhattacharya2019group} also included a stability rule for the agents, which prevents them from sending requests between two large clusters, i.e. clusters having sizes larger than $60\%$ of the maximum possible value.
Even though this stability rule was enforced during the experiment, the agents were generally more active in terms of sending requests. 
This higher activity caused instability and resulted in breaking of otherwise beneficial clusters in the game as the agents' goal was to reach the maximal clusters instead of obtaining points by forming clusters of size 6 or 9. 

We measure the behaviour of the players using measures of activity, risk averseness and rationality 
of the taken actions based on an agent-based model implemented in our previous study \cite{bhattacharya2019group}.
The activity corresponding to requesting an exchange of places with a nearest neighbour or accepting one of such requests, is measured as the rate of performing the action whenever it is allowed for the focal player. 
Hence we define the activity of a player $i$ as:
\begin{equation}
    a_i = \frac{N_i}{\mathcal{N}_i},
\end{equation}
where $N_i$ is the actual number of instances at which the focal player $i$ chose to interact with a neighbour and $\mathcal{N}_i$ is the total number of instances when the player $i$ had an option to interact. 
The quantity $a_i$ is separately measured for the requesting and accepting actions. 
For instance, a requesting activity of $1.0$ indicates that a player sent a request every time the player had the option. 
The notion of risk averseness is very similar to the activity but is measured as a function of the player's cluster size. 
The lower the value, the less a player is willing to risk losing the cluster size and eventually points the player already possesses. For a player $i$ with cluster $s\in\left[1,l\right]$, the risk averseness is quantified using the following ratio:
\begin{equation}
    a_i(s) = \frac{N_i(s)}{\mathcal{N}_i(s)},
\end{equation}
where $N_i(s)$ is the number of instances when the  player has a cluster size $s$ and interacts with a neighbour, and $\mathcal{N}_i(s)$ denotes the total number of instances when the player $i$ has a cluster size $s$ and is in a position to interact. 
Again, $a_i(s)$ is separately defined for requesting and accepting. 
We define the rationality of players in the following way. 
For the focal player $i$ that sends or accepts requests with respect to a neighbour $j$, we measure the quantity:
\begin{equation}
   U_j=\langle s(c_j) \rangle - s_j. 
\label{eq:eqn-uj}
\end{equation}
Here, $c_j$ is the colour of $j$, $\langle s(c_j) \rangle$ is average of the cluster sizes of the neighbours of $i$ having a colour $c_j$, that is, the same as that of $j$, and $s_j$ is the cluster size of $j$. 
Note that the information on all the cluster sizes accounted above is available to $i$ and not to $j$. 
The larger the difference between $\langle s(c_j) \rangle$ and $s_j$, the more is the expected gain for the neighbour $j$ if an exchange is facilitated by player $i$. 
Therefore, large positive values would indicate a cooperative strategy on the part of the player $i$. 
In a purely cooperative setup this quantity would reflect on rational choices and cognition of the players~\cite{bhattacharya2019group}.

In our previous experiment it was noticed that the players' activity in sending and accepting requests decreased the closer they were to the objective of forming maximal clusters. 
In the presence of a different point-based payoff function the players depicted a similar behavior when sending requests, but showed a more relaxed decision making when accepting incoming requests after achieving certain number of points (see Fig. \ref{fig_activity}). 
The less strict accepting behavior after reaching the required cluster size shows that the players understood the payoff function and adjusted their decision making after reaching the beneficial cluster sizes. 
This increase in one's activity also shows that the players trying to optimize their points can sacrifice their position for helping other smaller clusters, thus gaining more points if the other groups reach larger clusters. 
However, due to the limitations of the game and the obtained sample size, some situations rarely appeared during the games, which can be seen from the fluctuation caused by the lack of players with cluster size 8 in the setup B (Fig.  \ref{fig_activity}). 
Evaluating the individual activities of the players in each setup does not show a significant difference between the setups as a whole, suggesting that the players did not behave differently in terms of overall activity (see Fig. \ref{fig_activity}). 
However, the activity of the human subjects adapted towards the setup D as can be seen from the Fig. \ref{fig_activity}.

As the activity of the human subjects suggests a difference between the decision making in the purely human setup A and the hybrid setups B and C, we investigate the rationality of the decision making of the human players. 
This averaged rationality of the choices over the possible cluster sizes is illustrated in Fig. \ref{strategies}.
Individually the players' strategies are heterogeneous, even with some of the players resulting in an average strategy with both negative value in accepting and requesting (see Fig. \ref{strategies}). 
This suggests that during those particular setups those players sent their requests to near neighbours with higher cluster size than the average in the specific neighbourhood.
These decisions are not necessarily irrational, as the players can have information that is not obtained from the neighbourhood, but derived from the network structure and the previous moves the player has taken.
The autonomous agents are excluded from this analysis as their decision making is homogeneous due to the implementation.
However, the environment in the game can result in some variation in our measures for strategy as the agents make their decisions based on a probability function over the possible choices, which in small sample sizes can show fluctuations.

\section*{Discussion}
In this study we have focused on using human-agent experiments to get insight into complex dynamics of coordination, cooperation, and decision making in interactions between humans and autonomous agents mimicking human behaviour. While the simulations using only autonomous agents can result in better or close to optimal performance, the results can vary when combining agents and humans in cooperative decision-making setups, as demonstrated in recent studies using communication or video game setups \cite{hu2020other, carroll2019utility}. 
These studies have shown that models for autonomous agents tend to perform worse when paired with human decision making whilst outperforming purely human groups coordinating themselves, unless the models are specifically trained to perform well with human subjects \cite{carroll2019utility}.
However, the underlying reasons for the worse performance when cooperating with autonomous agents can stem from the lack of rule-based or uniform decision-making by the human players. 
In the case of comparing purely human groups to groups of agents, these type of differences can also stem from the higher operational capabilities of the agents in the current environment \cite{berner2019dota}.  
In the present study we experimented on hybrid human-agents collectives in a cooperative group-formation game to evaluate the performance of such groups with varying compositions and to investigate the effects of non-uniform payoff functions. 
These autonomous agents were modelled using the results from our previous study \cite{bhattacharya2019group} where the goal was purely collective, which is how to converge to a win-win situation in a limited number of rounds. 
As the objective of these agents would differ from the human objective, the questions were how and to what degree the agents modelled on outcomes from purely cooperative and uniformly rewarded payoff function could influence the dynamics and decision making of the human subjects. 
In addition, we have examined the differences in the outcomes from hybrid setups that were different in terms of the composition of humans and agents in the teams.
Our initial anticipation based on running simulations with varying types of agent models, including the one in the setups, was that the human-agent groups would facilitate more movement and thus achieve the individualistic payoff faster than groups consisting of humans with more prudent behaviour.

For better understanding of the above aspects, the design of the experiment was split into three different setups with varying compositions of autonomous agents in two of the experimental setups.
A control group for the agents was implemented with a setup of only autonomous agents using the same model as in these two hybrid setups.
The experimental design of the present study followed that of the previous experiments with the choice of network topology as a mesh with periodic boundary and additional small-world links to facilitate the cognition and movement of the human players. 
However, as achieving the maximum payoff did not require the formation of the maximum clusters, the given number of rounds was reduced from 21 to 15, giving each of the three colour groups five opportunities to send a request for swapping places with their neighbours. 
The choice of the reduced number of rounds was motivated by training agents with the same point-based payoff function and measuring the average completion time for the maximum payoff.
This preliminary simulation of possible outcomes reinforced our initial anticipation that the human-agent groups would be more effective.
The number of games was also limited by the length of the experimental session. 
Lengthening the session further could cause tiredness in the human players and thus affect their decision-making. 
It should be noted that this is a limitation imposed by the game design as the appearance of all possible states (i.e. combinations of different neighbourhoods) in sufficient numbers cannot be ensured during an experimental session.

First, in the case of human-human setup the game showed that the human subjects had comprehended the modified individualistic payoff function, even though the end result of each game in the setup was the formation of maximal groups as in the collective payoff games. 
The manifestation of this particular payoff was clearly seen in the human subjects' accepting actions as the fraction of accepted requests for each cluster size follows the objective cluster sizes instead of a declining slope like in the collective payoff games with a payoff based on the cluster size.
The measures for the strategies of the human players take this into account by aggregating the states of the game into combinations of three variables, i.e. cluster size of the focal player $s_i$, the chosen alter’s cluster size $s_j$ and the average cluster size of these neighbouring players sharing the colour of the chosen alter $\langle s(c_j)\rangle$.
The payoff function affected the players' strategy by making them more prone to accept incoming requests once their cluster size exceeded the required threshold. 
The same effect was not present in the request action, hinting that the human players split their strategy into achieving the maximum cluster (request) and helping other groups to form their clusters (accept). 
This imbalance of objectives was also evidenced in the experimental human-agent setups as the following. 
The agents modelled, by the players with the fully cooperative setup, performed well with the human subjects following a different payoff function, but when the agents played the game amongst themselves, the performance exceeded the performance in the hybrid setups. 
This suggests that the simultaneous presence of different strategies could be detrimental to the performance, thus proving our initial anticipation to be wrong, at least with the model the agents' were based on. 
It should be noted that the agents had a fixed strategy, while the human players did not as they had to adapt to the agents' strategies.

To sum up, our game had a payoff function that was partitioned into three types of benefits such that two of them were achievable from the performance at the player's team level (a small and a large threshold for player's own group size) and the third type that depended on the performance of the other teams. To achieve the third type of benefit, a team could be required to cooperate with other teams, which however also entailed some risk as its own cluster size could fall below the thresholds. 
Our previous experiment (incentivized on a collective goal) \cite{bhattacharya2019group} demonstrated that it is possible for all the teams to simultaneously achieve the maximum benefit if they all cooperated. Therefore, by using a setup with all human subjects we tested whether a different payoff function that made the game to appear less cooperative would modify the outcome of the game. 
Next, we focused on how the overall performance was modified with the inclusion of the agents and how the type of mixing (setup B versus setup C) would influence the outcome. Our experiments allow us to make two broad observations, first, that the overall performance is best when the teams are composed of solely humans (setup A) and second, that there is a hint that homogeneously composed human and agent teams (setup B) might lead to lowering of the performance in contrast to having separate teams of humans and agents (setup C). The plot for the average points reveals that quite fast increase is possible in the setups A and C. In cases when there is 
only human players in setup A and a single team totally composed of human players in setup C,  
would imply better control of the overall dynamics by the human teams. 
This could be because of the superior information processing and adaptation by the human subjects in comparison to the autonomous agents. 
Also, the behaviour captured by 
our parsimonious model could limit the performance of the autonomous agents, thus for better performance 
a more accurate behavioral model of cooperative human subjects might be needed.
Models taking into account human capabilities in terms of memory of previous states and approximating the information beyond the local neighbourhood could also provide more accurate depiction of human performance in the game of group formation.
As stated above human decisions are evidenced as a mixture of being risk averse and making cooperative moves. The human teams appear to flexibly adopt different strategies during the different stages of the game. Therefore, in either purely or somewhat lesser cooperative settings, providing more control to humans in hybrid systems (as in setup C) could help maximize the overall performance of the system. 
Alternatively, hybrid setups with lesser number of agent-agent interactions are expected to perform better. However, if the payoff is further skewed such that there is even lesser or no benefit for the teams to help each other, then it is plausible that cooperative behaviour of the agents could be instrumental in improving the overall performance of the whole system.

\bibliographystyle{unsrt}
\bibliography{refs}

\begin{thebibliography}{10}

\bibitem{kearns2006experimental}
Michael Kearns, Siddharth Suri, and Nick Montfort.
\newblock An experimental study of the coloring problem on human subject
  networks.
\newblock {\em Science}, 313(5788):824--827, 2006.

\bibitem{jennings2014human}
Nicholas~R Jennings, Luc Moreau, David Nicholson, Sarvapali Ramchurn, Stephen
  Roberts, Tom Rodden, and Alex Rogers.
\newblock Human-agent collectives.
\newblock {\em Communications of the ACM}, 57(12):80--88, 2014.

\bibitem{isern2010agents}
David Isern, David S{\'a}nchez, and Antonio Moreno.
\newblock Agents applied in health care: A review.
\newblock {\em International journal of medical informatics}, 79(3):145--166,
  2010.

\bibitem{corchado2008intelligent}
Juan~M Corchado, Javier Bajo, Yanira De~Paz, and Dante~I Tapia.
\newblock Intelligent environment for monitoring alzheimer patients, agent
  technology for health care.
\newblock {\em Decision Support Systems}, 44(2):382--396, 2008.

\bibitem{van2017domo}
Jenny Van~Doorn, Martin Mende, Stephanie~M Noble, John Hulland, Amy~L Ostrom,
  Dhruv Grewal, and J~Andrew Petersen.
\newblock Domo arigato mr. roboto: Emergence of automated social presence in
  organizational frontlines and customers’ service experiences.
\newblock {\em Journal of Service Research}, 20(1):43--58, 2017.

\bibitem{robu2013online}
Valentin Robu, Enrico~H Gerding, Sebastian Stein, David~C Parkes, Alex Rogers,
  and Nick~R Jennings.
\newblock An online mechanism for multi-unit demand and its application to
  plug-in hybrid electric vehicle charging.
\newblock {\em Journal of Artificial Intelligence Research}, 48:175--230, 2013.

\bibitem{bonabeau2002agent}
Eric Bonabeau.
\newblock Agent-based modeling: Methods and techniques for simulating human
  systems.
\newblock {\em Proceedings of the National Academy of Sciences}, 99(suppl
  3):7280--7287, 2002.

\bibitem{kahneman2003maps}
Daniel Kahneman.
\newblock Maps of bounded rationality: Psychology for behavioral economics.
\newblock {\em American economic review}, 93(5):1449--1475, 2003.

\bibitem{groom2007can}
Victoria Groom and Clifford Nass.
\newblock Can robots be teammates?: Benchmarks in human--robot teams.
\newblock {\em Interaction Studies}, 8(3):483--500, 2007.

\bibitem{bhattacharya2019group}
Kunal Bhattacharya, Tuomas Takko, Daniel Monsivais, and Kimmo Kaski.
\newblock Group formation on a small-world: experiment and modelling.
\newblock {\em Journal of the Royal Society Interface}, 16(156):20180814, 2019.

\bibitem{ishowo2019behavioural}
Fatimah Ishowo-Oloko, Jean-Fran{\c{c}}ois Bonnefon, Zakariyah Soroye, Jacob
  Crandall, Iyad Rahwan, and Talal Rahwan.
\newblock Behavioural evidence for a transparency--efficiency tradeoff in
  human--machine cooperation.
\newblock {\em Nature Machine Intelligence}, 1(11):517--521, 2019.

\bibitem{shirado2017locally}
Hirokazu Shirado and Nicholas~A Christakis.
\newblock Locally noisy autonomous agents improve global human coordination in
  network experiments.
\newblock {\em Nature}, 545(7654):370, 2017.

\bibitem{huang2018heterogeneous}
Keke Huang, Xiaofang Chen, Zhaofei Yu, Chunhua Yang, and Weihua Gui.
\newblock Heterogeneous cooperative belief for social dilemma in multi-agent
  system.
\newblock {\em Applied Mathematics and Computation}, 320:572--579, 2018.

\bibitem{huang2018incorporating}
Keke Huang, Zhen Wang, and Marko Jusup.
\newblock Incorporating latent constraints to enhance inference of network
  structure.
\newblock {\em IEEE Transactions on Network Science and Engineering},
  7(1):466--475, 2018.

\bibitem{perc2007cyclical}
Matja{\v{z}} Perc, Attila Szolnoki, and Gy{\"o}rgy Szab{\'o}.
\newblock Cyclical interactions with alliance-specific heterogeneous invasion
  rates.
\newblock {\em Physical Review E}, 75(5):052102, 2007.

\bibitem{kearns2012behavioral}
Michael Kearns, Stephen Judd, and Yevgeniy Vorobeychik.
\newblock Behavioral experiments on a network formation game.
\newblock In {\em Proceedings of the 13th ACM Conference on Electronic
  Commerce}, pages 690--704. ACM, 2012.

\bibitem{judd2010behavioral}
Stephen Judd, Michael Kearns, and Yevgeniy Vorobeychik.
\newblock Behavioral dynamics and influence in networked coloring and
  consensus.
\newblock {\em Proceedings of the National Academy of Sciences},
  107(34):14978--14982, 2010.

\bibitem{gale1962college}
David Gale and Lloyd~S Shapley.
\newblock College admissions and the stability of marriage.
\newblock {\em The American Mathematical Monthly}, 69(1):9--15, 1962.

\bibitem{laureti2003matching}
Paolo Laureti and Yi-Cheng Zhang.
\newblock Matching games with partial information.
\newblock {\em Physica A: Statistical Mechanics and its Applications},
  324(1-2):49--65, 2003.

\bibitem{baronchelli2010modeling}
Andrea Baronchelli, Tao Gong, Andrea Puglisi, and Vittorio Loreto.
\newblock Modeling the emergence of universality in color naming patterns.
\newblock {\em Proceedings of the National Academy of Sciences},
  107(6):2403--2407, 2010.

\bibitem{guazzini2015modeling}
Andrea Guazzini, Daniele Vilone, Camillo Donati, Annalisa Nardi, and Zoran
  Levnaji{\'c}.
\newblock Modeling crowdsourcing as collective problem solving.
\newblock {\em Scientific reports}, 5:16557, 2015.

\bibitem{centola2015spontaneous}
Damon Centola and Andrea Baronchelli.
\newblock The spontaneous emergence of conventions: An experimental study of
  cultural evolution.
\newblock {\em Proceedings of the National Academy of Sciences},
  112(7):1989--1994, 2015.

\bibitem{chauhan2018Schelling}
Ankit Chauhan, Pascal Lenzner, and Louise Molitor.
\newblock Schelling segregation with strategic agents.
\newblock In Xiaotie Deng, editor, {\em Algorithmic Game Theory}, pages
  137--149, Cham, 2018. Springer International Publishing.

\bibitem{elkind2019schelling}
Edith Elkind, Jiarui Gan, Ayumi Igarashi, Warut Suksompong, and Alexandros~A
  Voudouris.
\newblock Schelling games on graphs.
\newblock {\em arXiv preprint arXiv:1902.07937}, 2019.

\bibitem{schelling1971dynamic}
Thomas~C Schelling.
\newblock Dynamic models of segregation.
\newblock {\em Journal of mathematical sociology}, 1(2):143--186, 1971.

\bibitem{dreze1980hedonic}
Jacques~H Dreze and Joseph Greenberg.
\newblock Hedonic coalitions: Optimality and stability.
\newblock {\em Econometrica (pre-1986)}, 48(4):987, 1980.

\bibitem{aziz2016hedonic}
Haris Aziz and Rahul Savani.
\newblock Hedonic games.
\newblock In Felix Brandt and Ariel~D Procaccia, editors, {\em Handbook of
  computational social choice}, chapter~15, pages 356--377. Cambridge
  University Press, New York, 2016.

\bibitem{barrett2007cooperate}
Scott Barrett et~al.
\newblock {\em Why cooperate?: the incentive to supply global public goods}.
\newblock Oxford University Press on Demand, 2007.

\bibitem{takko2019study}
Tuomas Takko.
\newblock Study on modelling human behavior in cooperative games.
\newblock Master's thesis, Aalto University, 2019.

\bibitem{otree}
Daniel~L. Chen, Martin Schonger, and Chris Wickens.
\newblock Otree - an open-source platform for laboratory, online and field
  experiments.
\newblock {\em Journal of Behavioral and Experimental Finance}, 9:88--97, 2016.

\bibitem{crandall2018cooperating}
Jacob~W Crandall, Mayada Oudah, Fatimah Ishowo-Oloko, Sherief Abdallah,
  Jean-Fran{\c{c}}ois Bonnefon, Manuel Cebrian, Azim Shariff, Michael~A
  Goodrich, Iyad Rahwan, et~al.
\newblock Cooperating with machines.
\newblock {\em Nature communications}, 9(1):233, 2018.

\bibitem{hu2020other}
Hengyuan Hu, Adam Lerer, Alex Peysakhovich, and Jakob Foerster.
\newblock " other-play" for zero-shot coordination.
\newblock {\em arXiv preprint arXiv:2003.02979}, 2020.

\bibitem{carroll2019utility}
Micah Carroll, Rohin Shah, Mark~K Ho, Tom Griffiths, Sanjit Seshia, Pieter
  Abbeel, and Anca Dragan.
\newblock On the utility of learning about humans for human-ai coordination.
\newblock In {\em Advances in Neural Information Processing Systems}, pages
  5174--5185, 2019.

\bibitem{berner2019dota}
Christopher Berner, Greg Brockman, Brooke Chan, Vicki Cheung, Przemys{\l}aw
  D{\k{e}}biak, Christy Dennison, David Farhi, Quirin Fischer, Shariq Hashme,
  Chris Hesse, et~al.
\newblock Dota 2 with large scale deep reinforcement learning.
\newblock {\em arXiv preprint arXiv:1912.06680}, 2019.

\end{thebibliography}

\section*{Acknowledgments}
All the authors acknowledge the support from EU HORIZON2020 FET  Open  RIA  project (IBSEN) No. 662725, SoBigData: Social Mining and Big Data Ecosystem, Grant agreement No. 654024, and  SoBigData++: European Integrated Infrastructure for Social Mining and Big Data Analytics, Grant agreement ID: 871042, (http://www.sobigdata.eu).

\section*{Author contributions}
T.T., K.B. and D.M. designed the study and organized the experimental session.
T.T. analyzed the results and created the figures.
T.T. and K.B. were responsible to write the original draft.
All authors contributed in discussion, writing and reviewing of the paper.
\section*{Competing interests}
The authors declare no competing interests.
\end{document}